\documentclass[aps,prl,reprint,superscriptaddress,10pt,a4paper,preprintnumbers,nofootinbib]{revtex4-1}
\usepackage{amsmath,amssymb,latexsym}
\usepackage{hyperref}
\usepackage{bm}
\usepackage{graphicx}
\usepackage{hepunits}
\usepackage[svgnames]{xcolor}
\usepackage{lipsum}

\renewcommand{\Re}{\mathrm{Re}}

\allowdisplaybreaks

\begin{document}

\title{Efficient computation of two-loop amplitudes for Higgs boson pair production }

\author{Guoxing Wang}
\affiliation{Zhejiang Institute of Modern Physics, Department of Physics, Zhejiang University, Hangzhou 310027, China}
\author{Yuxuan Wang}
\affiliation{School of Physics and State Key Laboratory of Nuclear Physics and Technology, Peking University, Beijing 100871, China}
\author{Xiaofeng Xu}
\email{pkuxxf@gmail.com}
\affiliation{Institut f\"ur Theoretische Physik, Universit\"at Bern, Sidlerstrasse 5, CH-3012 Bern, Switzerland}
\author{Yongqi Xu}
\email{xuyongqi@pku.edu.cn}
\affiliation{School of Physics and State Key Laboratory of Nuclear Physics and Technology, Peking University, Beijing 100871, China}
\author{Li Lin Yang}
\email{yanglilin@zju.edu.cn}
\affiliation{Zhejiang Institute of Modern Physics, Department of Physics, Zhejiang University, Hangzhou 310027, China}

\begin{abstract}
We present a precise and efficient computation of the two-loop amplitudes entering the Higgs boson pair production process via gluon fusion. Our approach is based on the small-Higgs-mass expansion while keeping the full dependence on the top quark mass and other kinematic invariants. We compare our results to the up-to-date predictions based on a combination of sector decomposition and high-energy expansion. We find that our method provides precision numeric predictions in the entire phase space, while at the same time is highly efficient as the computation can be easily performed on a normal desktop or laptop computer. Our method is valuable for practical phenomenological studies of the Higgs boson pair production process, and can also be applied to other similar processes.
\end{abstract}


\maketitle

\section{Introduction}

Higgs boson pair production via gluon fusion is one of the most important scattering processes yet to be discovered at the Large Hadron Collider (LHC). It provides information on the Higgs trilinear self-coupling (which is an important parameter in the Higgs potential, relevant for electroweak symmetry breaking and vacuum stability), or more generically, on the  Wilson coefficients of the relevant effective operators encoding physics at higher scales beyond the standard model (SM) \cite{Goertz:2014qta, Azatov:2015oxa, Cao:2015oaa, DiVita:2017eyz}. Hence, precision theoretical predictions for this process (both within the SM and within the effective field theory of beyond-the-SM physics) are highly important for a future extraction of these crucial information from experimental measurements of the differential cross sections.

At the leading order (LO) in the SM couplings, the process goes mainly through a top quark loop, with a small contribution from a bottom quark loop. The relevant LO Feynman diagrams can be categorized into two types: the triangle diagrams which are sensitive to the Higgs trilinear self-coupling, and the box diagrams which are sensitive to the top quark Yukawa coupling. In the SM, these two types of diagrams numerically cancel against each other, giving a rather small cross section \cite{Glover:1987nx, Plehn:1996wb}.
New physics beyond the SM will modify these couplings and may spoil this accidental cancellation. In that case the production rate will be significantly enhanced. An early observation of this process at the LHC is then a clear evidence of new physics lying not too far above the electroweak scale.

Due to the importance of the Higgs boson pair production process, its precision theoretical prediction has been an active research area in the past few years. The LO results have been available since about 30 years ago \cite{Glover:1987nx, Plehn:1996wb}. The calculation of the next-to-leading order (NLO) corrections in quantum chromodynamics was quite challenging, due to the appearance of difficult two-loop Feynman integrals whose analytic expressions are still unknown so far.
Because of that, people have attempted different methods to approximately calculate these two-loop integrals. The most straightforward way is to compute the integrals numerically using the method of sector decomposition \cite{Binoth:2000ps, Bogner:2007cr, Heinrich:2008si}. This approach is rather resource-consuming and has only become possible recently with the help of modern computer clusters \cite{Borowka:2016ehy, Borowka:2016ypz, Heinrich:2017kxx, Buchalla:2018yce}. Besides the automated computation using sector decomposition, Ref.~\cite{Baglio:2020ini} calculated these integrals by carefully manipulating the Feynman-parameterized integrals supplemented by numeric integration and extrapolation.

Apart from the purely numerical method, it is also possible to approximate the integrals by working with a series expansion in terms of one (or more) small parameter(s) in certain kinematic limits. The simplest one is the $1/m_t$ expansion \cite{Dawson:1998py, deFlorian:2013uza, deFlorian:2013jea, deFlorian:2016uhr, Grigo:2013rya, Grigo:2015dia, Degrassi:2016vss}, whose leading power is dubbed the Higgs effective field theory (HEFT) in the literature. People have also studied the small transverse momentum ($p_T$) expansion \cite{Bonciani:2018omm} and the high-energy expansion \cite{Davies:2018ood, Davies:2018qvx, Mishima:2018olh}. All these expansions work in a specific kinematic region, and cannot be directly applied to the whole phase space. Recently, the high-energy expansion has been combined with the results of sector decomposition via the Pad\'e approximation, aiming at a good description of the kinematic distribution from low to high energies \cite{Davies:2019dfy}. Later on we will compare our results to the last one in more detail.

In \cite{Xu:2018eos}, some of the authors of this Letter proposed a novel method to calculate loop integrals with a heavy top quark loop and lighter external particles such as the Higgs boson and the vector bosons in the SM. In this Letter, we apply this method to calculate the complete NLO virtual corrections to the Higgs boson pair production process via gluon fusion. We show that our method provides accurate numeric predictions for the two-loop amplitudes in the entire physical phase space. At the same time, the numeric performance of our method is rather efficient, and can be accomplished on a desktop computer or even a laptop. The precision and efficiency of our approach makes it convenient for phenomenological studies. Our method can also be implemented for other $2 \to 2$ processes with a heavy quark loop, such as $gg \to ZZ$, $gg \to ZH$ and $gg \to Hj$.

\section{Calculation of the amplitudes}

We consider the partonic process $g^{\mu}_{a}(p_1)+g^{\nu}_{b}(p_2) \to H(p_3)+H(p_4)$, where $a$ and $b$ are color indices. The amplitude can be written as
\begin{equation}
\mathcal{M}^{\mu\nu}_{ab} = \frac{G_F}{\sqrt{2}} \frac{\alpha_s}{2\pi} \, \hat{s} \, \delta_{ab} \, \Big[ A_1^{\mu\nu} F_1 + A_2^{\mu\nu} F_2 \Big] \, ,
\label{eq:amp}
\end{equation}
where $G_F$ is the Fermi constant. The Mandelstam variables are defined as $\hat{s}=(p_1+p_2)^2$,  $\hat{t}_1=(p_1-p_3)^2-m_H^2$ and $\hat{u}_1=(p_2-p_3)^2-m_H^2$, which satisfy $\hat{s} + \hat{t}_1 + \hat{u}_1 = 0$. The tensor structures in Eq.~\eqref{eq:amp} are given by
\begin{align}
A_1^{\mu\nu} &= g^{\mu\nu} - \frac{2p_1^\nu \, p_2^\mu}{\hat{s}} \, , \nonumber
\\
A_2^{\mu\nu} &= g^{\mu\nu} + \frac{2 m_H^2 p_1^\nu p_2^\mu + 2 \hat{s} p_3^\mu p_3^\nu + 2 \, \hat{u}_1 p_1^\nu p_3^\mu + 2 \, \hat{t}_1 p_2^\mu p_3^\nu}{p_T^2 \hat{s}}\, ,
\label{eq:proj}
\end{align}
where $p_T$ denotes the transverse momentum of the Higgs boson with respect to the beam axis and can be expressed in terms of the Mandelstam variables as $p_T^2 = \hat{t}_1 \hat{u}_1 / \hat{s} - m_H^2$.
The corresponding form factors $F_{1,2}$ are functions of $\hat{s}$, $\hat{t}_1$, $m_t^2$ and $m_H^2$, and can be expressed as power series in the strong coupling $\alpha_s$:
\begin{align}
F_i(\hat{s},\hat{t}_1,m_t^2,m_H^2) = F^{(0)}_i + \frac{\alpha_s}{\pi} F^{(1)}_i + \mathcal{O}(\alpha_s^2) \, .
\end{align}
The NLO terms $F^{(1)}_i$ receive contributions from two-loop diagrams with a top quark loop and virtual gluon exchanges, which are the main concern of this work.

The two-loop Feynman integrals contain both ultraviolet (UV) and infrared (IR) divergences. The UV divergences are removed by renormalization, for that we adopt the on-shell scheme for $m_t$ and the five-flavor $\overline{\text{MS}}$ scheme for $\alpha_s$. The IR divergences will be eventually cancelled by the real-emission contributions and the renormalization of the parton distribution functions (PDFs). We follow the dipole subtraction method \cite{Catani:1998bh} to remove the IR divergences from the two-loop amplitudes, and define the finite part of the NLO form factors as \cite{Davies:2019dfy}
\begin{align}
\tilde{F}_{i}^{(1)} = F^{(1),\text{ren}}_{i} - K_g^{(1)} F_{i}^{(0)} - \beta_0 F^{(0)}_{i} \log\left(\frac{\mu^2}{-\hat{s}}\right) ,
\end{align}
where $F^{(1),\text{ren}}_{i}$ denote the UV-renormalized form factors, $\beta_0=11 C_A/12 - T_F N_l/3$ with $N_l = 5$, and $K_g^{(1)}$ is given by
\begin{align}
K_g^{(1)} &= - \left(\frac{\mu^2}{-\hat{s}}\right)^\epsilon \frac{e^{\epsilon\gamma_E}}{2\Gamma(1-\epsilon)}
\left( \frac{C_A}{\epsilon^2} + \frac{2}{\epsilon} \beta_0 \right) .
\label{eq::IRsub}
\end{align}
Here and below, one should remember that the physical branch is chosen by $\hat{s} \to \hat{s} + i\delta$.

In order to describe the contributions of the NLO form factors to the partonic cross section, we follow the notation of Refs.~\cite{Davies:2019dfy} and define the finite part of the two-loop virtual corrections as
\begin{equation}
\mathcal{V}_{\text{fin}} = \frac{1}{16\pi^2} \frac{G_F^2 \hat{s}^2}{64}
\big( C_0 + C_1 \big) \, ,
\label{eq::Vfin}
\end{equation}
with
\begin{align}
C_0 &= \left[ \big|F_1^{(0)}\big|^2 + \big|F_2^{(0)}\big|^2 \right]
C_A \left(\pi^2 - \log^2\frac{\mu^2}{\hat{s}} \right) , \nonumber
\\
C_1 &= 4 \, \Re \left[ F_1^{(0)} \tilde{F}_1^{(1)*} + F_2^{(0)} \tilde{F}_2^{(1)*} \right] .
\end{align}
The quantity $\mathcal{V}_{\text{fin}}$ involves various ingredients, among which the one-loop diagrams, the two-loop one-particle reducible diagrams, and the two-loop triangle diagrams can be calculated analytically \cite{Glover:1987nx, Plehn:1996wb, Degrassi:2016vss, Harlander:2005rq, Anastasiou:2006hc, Aglietti:2006tp}. The only difficult part is the two-loop one-particle irreducible box diagrams. We denote those contributions as $F_{i, \text{box}}^{(1)}$, and employ the small-Higgs-mass expansion proposed in \cite{Xu:2018eos} for their calculation.

We generate the two-loop virtual diagrams using \texttt{FeynArts} \cite{Hahn:2000kx}, and manipulate the resulting amplitudes with \texttt{FeynCalc} \cite{Shtabovenko:2016sxi} and \texttt{FORM} \cite{Vermaseren:2000nd}. The form factors are then expressed in terms of linear combinations of scalar integrals, which are functions of $\hat{s}$, $\hat{t}_1$, $m_t^2$ and $m_H^2$. These scalar integrals are rather complicated, and cannot be evaluated analytically using existing methods. In fact, even their reduction to master integrals is highly non-trivial \cite{Borowka:2016ehy, Borowka:2016ypz}. To deal with them, we observe that these integrals, and hence the form factors, can be expanded as Taylor series in $m_H^2$:
\begin{equation}
F_{i, \text{box}}^{(1)}(\hat{s},\hat{t}_1,m_t^2,m_H^2) = \sum_{n=0}^\infty  \frac{(m_H^2)^n}{n!} \left[ \frac{\partial^n F_{i, \text{box}}^{(1)}}{\partial (m_H^2)^n} \right]_{m_H^2=0} .
\end{equation}
When acting on the loop integrals, the partial derivative operator can be exchanged with the integration, which then acts on the integrands as
\begin{align}
\partial_{m_H^2} = \frac{ \hat{u}_1 p_1^\mu + \hat{t}_1 p_2^\mu + \hat{s} p_3^\mu}{2 m_H^2 \hat{s} - 2 \hat{t}_1 \hat{u}_1} \, \partial_{p_3^\mu} \, .
\label{eq:derivative}
\end{align}

\begin{figure}[t!]
\centering
\includegraphics[width=8.5cm,height=3.5cm]{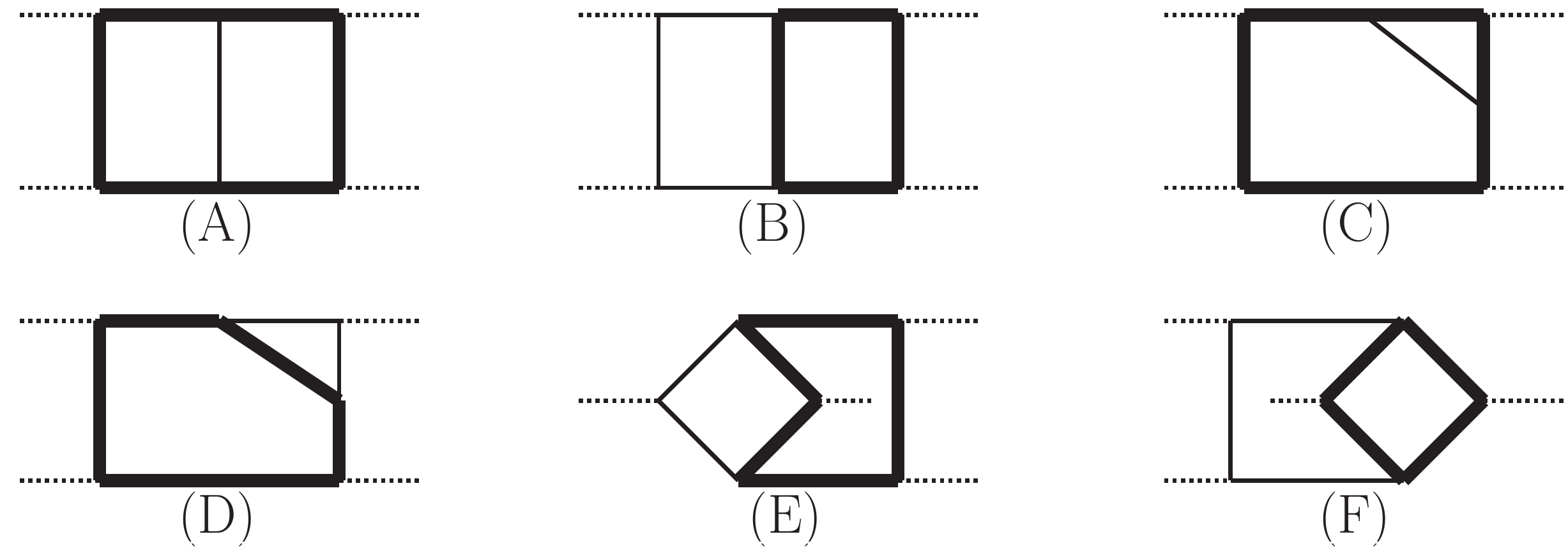}
\vspace{-1ex}
\caption{Integral topologies relevant for $F_{i,\text{box}}^{(1)}$. The thick lines represent massive top quark propagators, while the internal thin lines represent massless gluon propagators. All external lines are massless after performing the small-mass expansion.}
\label{fig:topology}
\end{figure}

After expansion, we need to deal with two-loop box diagrams with massless external legs. They can be categorized into 6 integral topologies shown in Fig.~\ref{fig:topology}.
We use program packages \texttt{FIRE6} \cite{Smirnov:2014hma, Smirnov:2019qkx} and \texttt{LiteRed} \cite{Lee:2012cn} to perform the integration-by-parts (IBP) reduction of these loop integrals.
There are 29, 32, 37, 27, 54, 36 master integrals in topologies A--F, respectively.
The first five topologies have been discussed in \cite{Caron-Huot:2014lda, Becchetti:2017abb, Xu:2018eos}. They can be solved using the method of canonical differential equations \cite{Henn:2013pwa} in terms of iterated integrals \cite{Chen:1977oja}. Most of these iterated integrals can be expressed as generalized polylogarithms (GPLs) \cite{Goncharov:1998kja}. We evaluate these GPLs using in-house routines and the program package \texttt{handyG} \cite{Naterop:2019xaf}. The remaining iterated integrals are expressed as one-fold integrals over GPLs of transcendental weight 2.

Topology F involves elliptic Feynman integrals and is more complicated. We have constructed a basis of the master integrals such that the $\epsilon^0$ part of their differential equations are as simple as possible, where $\epsilon$ is the dimensional regulator. The solutions of these differential equations involve iterated integrals over elliptic integrals of the first and second kinds, which we perform numerically. These integrals turn out to be the most time-consuming part of the numeric evaluation of the amplitudes. It is possible to speed up the computation using methods of, e.g., Refs.~\cite{Bogner:2019vhn, Walden:2020odh}.

\section{Numeric results and discussions}

In this section, we present some numeric results for the finite part of the two-loop amplitude, i.e., $\mathcal{V}_{\text{fin}}$ defined in Eq.~\eqref{eq::Vfin}, computed using our method. While these results are not new, our aim is to demonstrate the precision and efficiency that can be achieved from the small-mass expansion. We will compare our results with those computed using sector decomposition in \cite{Borowka:2016ehy, Borowka:2016ypz, Heinrich:2017kxx} (optionally supplemented by the Pad\'e-improved high-energy expansion \cite{Davies:2018ood, Davies:2018qvx, Mishima:2018olh, Davies:2019dfy}). For that purpose we choose the same input parameters as in \cite{Davies:2019dfy}: the Higgs mass $m_H=\SI{125}{\GeV}$,  the top quark mass $m_t=\SI{173}{\GeV}$, and the renormalization scale $\mu=\sqrt{\hat{s}}/2$.

\begin{figure}[t!]
\leftskip -15pt
\includegraphics[width=1.12\linewidth]{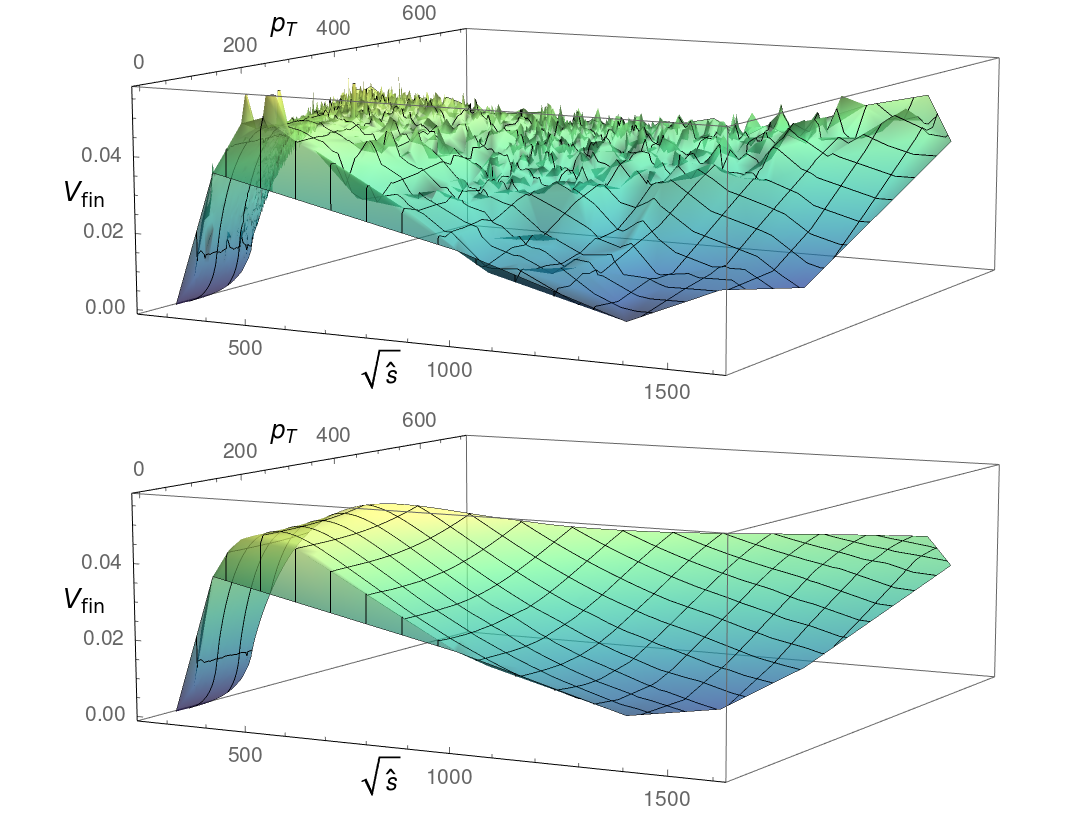}
\vspace{-5ex}
\caption{The finite part $\mathcal{V}_{\text{fin}}$ of the two-loop amplitude as a function of $\sqrt{\hat{s}}$ and $p_T$. The upper plot uses the results from the grid file of \cite{Davies:2019dfy}, while the lower one uses our results with the small-mass expansion up to $\mathcal{O}(m_H^4)$. See the text for how the plots are generated.}
\label{fig:expansion_vs_num_3D}
\end{figure}

Our approach is a mixture of analytic and numeric methods. Its main advantage lies in the fact that it is much more efficient than the purely numeric integration with sector decomposition, yet provides even better precision in a major portion of the whole phase space. To see that, we show $\mathcal{V}_{\text{fin}}$ as a function of $\sqrt{\hat{s}}$ and $p_T$ in Fig.~\ref{fig:expansion_vs_num_3D}. In order to generate the surfaces, we plot the results of $\mathcal{V}_{\text{fin}}$ at 6320 phase-space points corresponding to the grid of \cite{Davies:2019dfy}, and then simply join them without any interpolation or fitting procedure.

The upper plot in Fig.~\ref{fig:expansion_vs_num_3D} uses the values of $\mathcal{V}_{\text{fin}}$ taken from the grid file of \cite{Davies:2019dfy}, which were computed using sector decomposition. It can be seen that the surface has a lot of spikes and is far from smooth. This is due to the uncertainties of numeric integrations, which are most severe when $\sqrt{\hat{s}}$ is above the $2m_t$ threshold. The integration uncertainties can in principle be reduced with more sampling points in the (quasi-)Monte Carlo methods. However, this requires much more computational resources which prevents its viability. To cure this spiky behavior, Ref.~\cite{Davies:2019dfy} has created an interpolation code which applies a Clough-Tocher interpolator resulting in a smooth distribution. This, however, does not reduce the intrinsic uncertainties of the results.

The lower plot in Fig.~\ref{fig:expansion_vs_num_3D}, on the other hand, uses our results of $\mathcal{V}_{\text{fin}}$ computed with the small-mass expansion up to $\mathcal{O}(m_H^4)$. One can see that the double distribution is rather smooth in the whole phase-space region probed by the 6320 points.  We emphasize again that the surface is generated without any interpolation or fitting procedure. This demonstrates the high precision and stability of our method, which is crucial for practical phenomenological applications. Due to the high efficiency of our method, it is straightforward to generate more phase-space points on a modern laptop computer. This helps to better describe the shape of the double distribution, which can then be easily integrated over without worrying too much about the interpolation.

\begin{figure}[t!]
\centering
\includegraphics[width=1.0\linewidth]{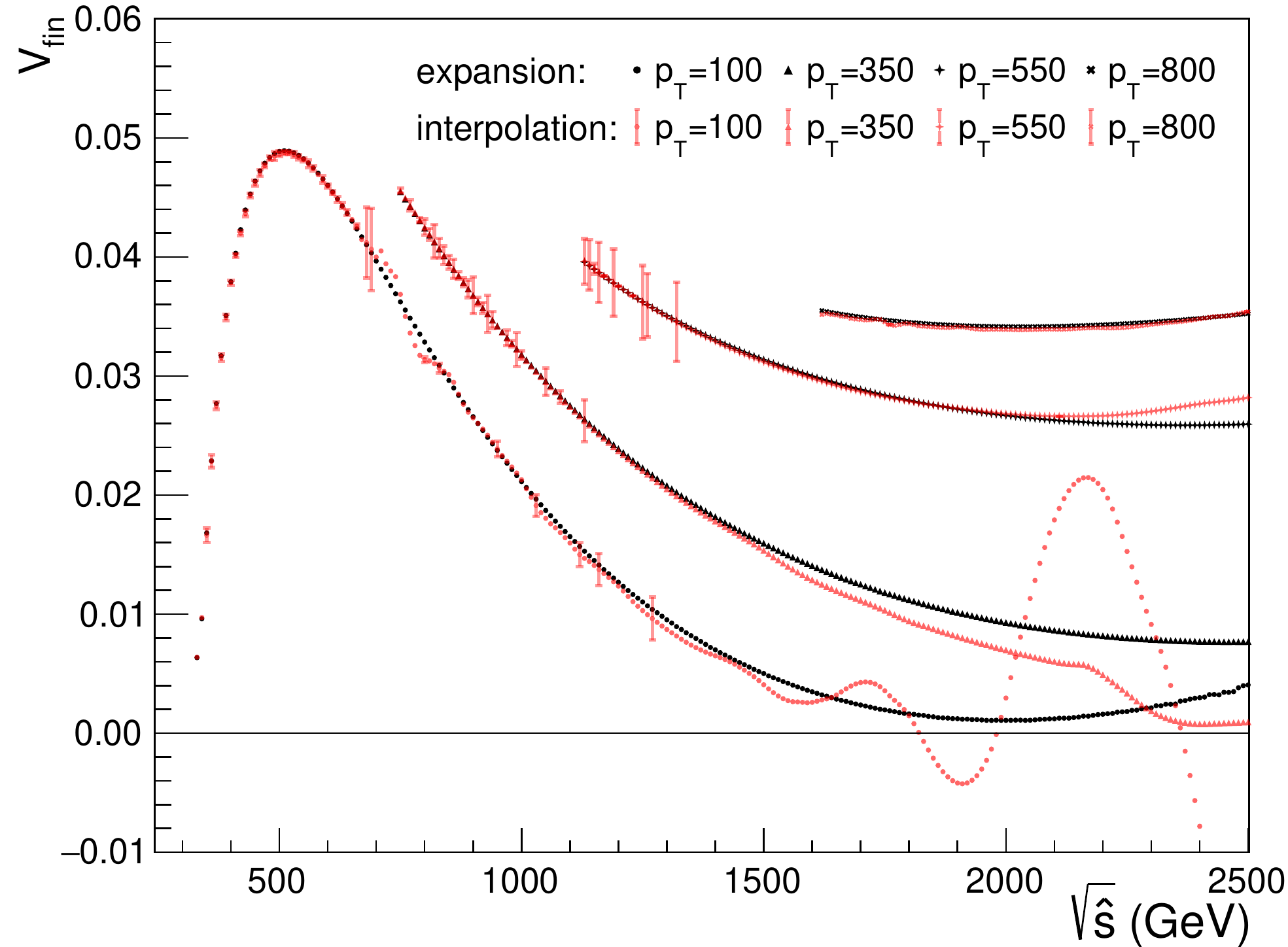}
\vspace{-5ex}
\caption{The finite part $\mathcal{V}_{\text{fin}}$ of the two-loop amplitude as a function of $\sqrt{\hat{s}}$ with fixed values of $p_T$. The black curves are obtained from our results with the small-mass expansion up to $\mathcal{O}(m_H^4)$, while the red ones are obtained using the interpolation code associated with \cite{Davies:2019dfy}.}
\label{fig:diff_pt2}
\end{figure}

The 3-D plots in Fig.~\ref{fig:expansion_vs_num_3D} only give a qualitative impression of the two approaches. To make a more quantitative comparison, in Fig.~\ref{fig:diff_pt2}, we present $\mathcal{V}_{\text{fin}}$ as a function of $\sqrt{\hat{s}}$ with fixed values of $p_T$. The black curves are obtained from our results with the small-mass expansion up to $\mathcal{O}(m_H^4)$, while the red ones are obtained using the interpolation code associated with \cite{Davies:2019dfy} (with the grid file containing results from the Pad\'e-improved high-energy expansion). The error bars of the interpolated results are calculated following the approach outlined in \cite{Davies:2019dfy}. We note that when no error bar is attached to a given point, it does \emph{not} necessarily mean that the result is accurate, but rather means that the result is not supported by nearby grid points and might not be reliable. The uncertainties of our results come mainly from the missing higher order terms at $\mathcal{O}(m_H^6)$, which would not be visible on the plot.
Overall, we find that our results are consistent with the interpolated ones wherever the interpolation is reliable (i.e., where there are enough grid points to support the interpolation and the associated numeric uncertainties are small). Also, our curves are smooth in all cases. These observations lend us confidence that the small-mass expansion provides a rather good approximation to the exact result across the whole phase space.

In Fig.~\ref{fig:diff_pt2}, one can also see that there are discrepancies between the two results in certain phase-space regions, which we now discuss.
In the small $p_T$ region (represented by the $p_T=\SI{100}{\GeV}$ curves), the high-energy expansion does not apply and the interpolated results are fully controlled by numeric integrations with sector decomposition. Since the numeric integration is time-consuming, the number of grid points are not sufficient to support the whole range of $\sqrt{\hat{s}}$ (here only those points with error bars are supported). Also, numeric integrations lead to large uncertainties at moderate and high $\sqrt{\hat{s}}$, which can be clearly seen in the plot. These two factors result in the wiggly behavior of the $p_T=\SI{100}{\GeV}$ curve for $\sqrt{\hat{s}} > \SI{700}{\GeV}$. On the other hand, our result remains smooth and is much more reliable in this region.

As $p_T$ increases, the Pad\'e-improved high-energy expansion comes into play. This helps to make the curve smoother, albeit that the uncertainties in the region of moderate $\sqrt{\hat{s}}$ are still large. The deviation in the tail (high $\sqrt{\hat{s}}$) region is due to the lack of grid points there to support the interpolation. Note that the interpolation in the tail region can in principle be improved by adding more data from high-energy expansion. Finally, at high $p_T$ (represented by the $p_T=\SI{800}{\GeV}$ curves), the two results almost completely coincide with each other, although the interpolated curve still has some small wiggles.

Given the high precision of our results in the whole phase space, we now turn to discuss the efficiency of our approach compared to that based on sector decomposition. Ref.~\cite{Borowka:2016ehy} quoted their performance of 4680 GPGPU hours (using NVIDIA Tesla K20X GPUs) for 665 phase-space points, giving an average of about 7 GPGPU hours per phase-space point. On the contrary, we only need about 10 seconds for one phase-space point using one CPU core on a desktop or even laptop computer. Note that a majority of the time in our approach is spent on the integration over elliptic functions, which may be further reduced using methods similar to those proposed in, e.g., \cite{Bogner:2019vhn, Walden:2020odh}. The high efficiency of our method makes it valuable for phenomenological analyses, in particular, for studies involving different values of $m_t$ and $m_H$.

\begin{figure}[t!]
\centering
\includegraphics[width=1.0\linewidth]{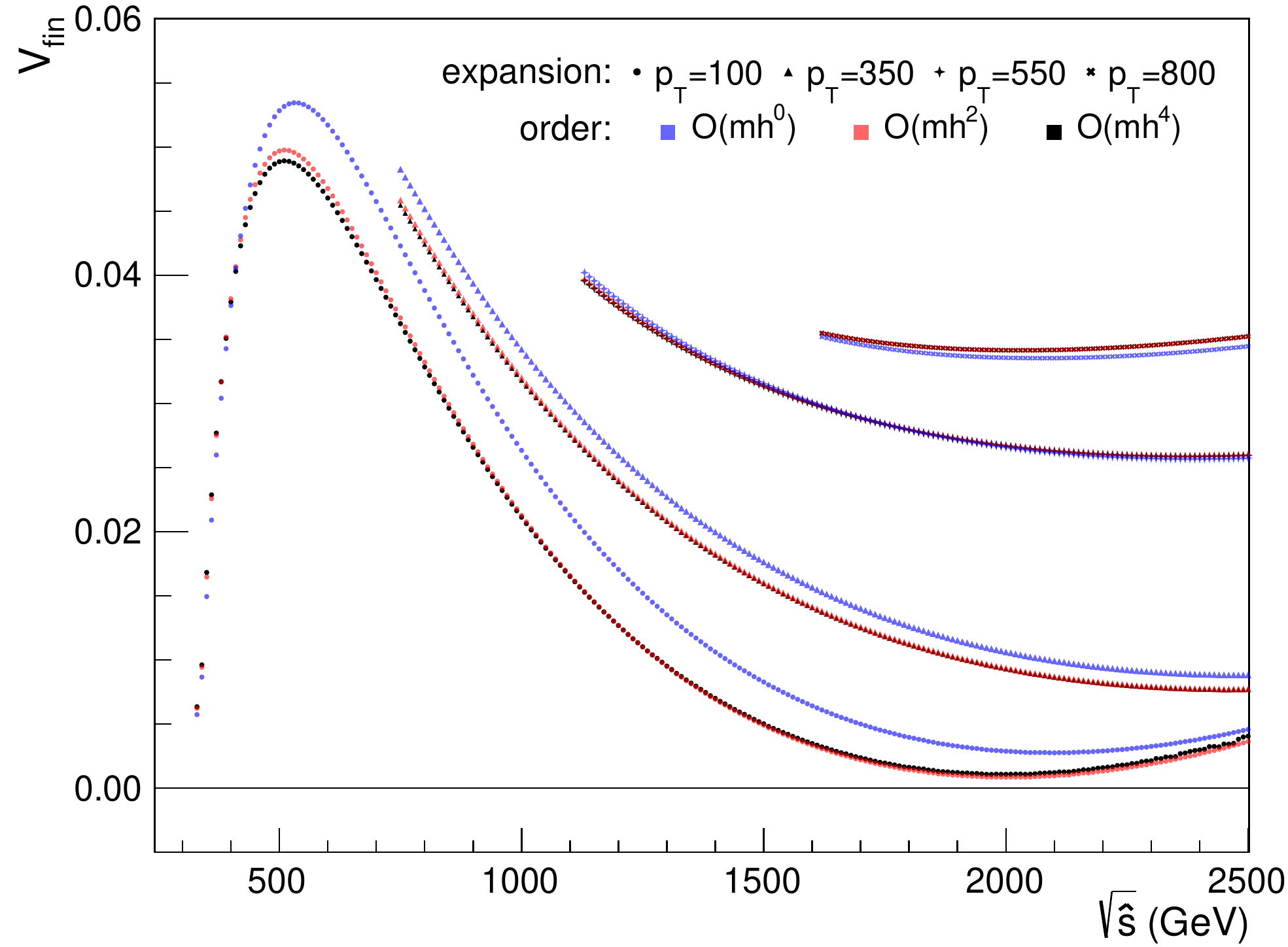}
\vspace{-5ex}
\caption{$\mathcal{V}_{\text{fin}}$ as a function of $\sqrt{\hat{s}}$ computed with the small-mass expansion up to $\mathcal{O}(m_H^0)$(blue), $\mathcal{O}(m_H^2)$(red), and $\mathcal{O}(m_H^4)$(black) for several values of $p_T$.}
\label{fig:convergence}
\end{figure}

In all previous numerics, we have truncated the small-mass expansion at $\mathcal{O}(m_H^4)$. It is now time to present the convergence of the expansion to justify this choice. In Fig.~\ref{fig:convergence} we plot $\mathcal{V}_{\text{fin}}$ computed using the small-mass expansion up to $\mathcal{O}(m_H^0)$, $\mathcal{O}(m_H^2)$ and $\mathcal{O}(m_H^4)$. We show $\mathcal{V}_{\text{fin}}$ as a function of $\sqrt{\hat{s}}$ for several representative values of $p_T$. We find that for large $p_T$, the $\mathcal{O}(m_H^2)$ corrections are already small, and the $\mathcal{O}(m_H^4)$ corrections are negligible. At small $p_T$, the $\mathcal{O}(m_H^2)$ effects can be significant, but the $\mathcal{O}(m_H^4)$ effects remain small. Therefore, we conclude that keeping the terms in the expansion up to $\mathcal{O}(m_H^4)$ is sufficient for most cases. Even if one finds that this is not enough, it is straightforward to add the $\mathcal{O}(m_H^6)$ terms, which does not introduce much computational overhead since one does not need to calculate new integrals beyond those already present at $\mathcal{O}(m_H^4)$.

\section{Conclusion}

In this Letter, we present an efficient method to calculate the two-loop amplitudes for Higgs boson pair production via gluon fusion. Our method is based on a series expansion in terms of the Higgs boson mass $m_H$. After expansion, the remaining integrals can be calculated with a combination of analytic and numeric methods.

The main advantage of our approach lies in two aspects: precision and efficiency. To demonstrate these, we compare our results with the most up-to-date calculation which combines the results from sector decomposition and those from Pad\'e-improved high-energy expansion. While in principle the method of sector decomposition can lead to the exact answer for the relevant integrals, in practice it is highly challenging to reduce the uncertainty of the Monte-Carlo integration, especially in the moderate and high energy regions. On the other hand, the integration uncertainty in our approach is negligible, and the uncertainty due to the expansion is under-control and can be easily reduced by including higher power terms in $m_H^2$. This leads to precision numeric results across the whole phase-space, as can be seen by the smooth behavior of kinematic distributions and the excellent agreement between our predictions and those from other approaches. Our method also gives rather good numeric performance, such that the computation can be carried out by a normal desktop or laptop computer. The precision and efficiency of our method is highly valuable for practical phenomenological analyses.

Our calculation can be easily extended to the cases with dimension-6 effective operators describing new physics beyond the SM, as the required two-loop integrals are the same. Our method can also be applied to Higgs boson pair production in a concrete new physics model, where new heavy particles such as the top quark partners may enter the loop. Finally, our method can be implemented for other $2 \to 2$ processes with a heavy quark loop, such as $gg \to ZZ$, $gg \to ZH$ and $gg \to Hj$, which we leave for future investigations.

\vspace{2ex}

\begin{acknowledgments}
This work was supported in part by the National Natural Science Foundation of China under Grant No. 11975030 and 11635001. The research of X. Xu was supported in part by the Swiss National Science Foundation (SNF) under Grant No. $200020\_182038$.
\end{acknowledgments}

\bibliography{references_inspire}

\end{document}